\documentstyle[epsfig,twoside,fleqn,espcrc2]{article}

\newcommand{\eq}{\begin{equation}}
\newcommand{\en}{\end{equation}}
\newcommand{\eqa}{\begin{eqnarray}}
\newcommand{\ena}{\end{eqnarray}}

\newcommand{\NP}[1]{Nucl.\ Phys.\ {\bf #1}}

\def\cj{{\cal C}^{(j)}}
\def\vx{{\vec x}}

\def\um{{1\over 2}}

\newcommand{\AmS}{{\protect\the\textfont2
  A\kern-.1667em\lower.5ex\hbox{M}\kern-.125emS}}

\title{Effective actions for finite temperature Lattice Gauge Theories.}

\author{M. Bill\'{o}\address{Nordita, Blegdamsvej 17, Copenhagen \O, Denmark},
M. Caselle\thanks{Speaker at the conference}$^{\rm b}$, 
A.D'Adda\address{
Dip. di Fisica Teorica, Universit\`a di Torino,
Via P. Giuria 1, 10125 Torino, Italy}
and S. Panzeri\address{SISSA,Via Beirut 2-4, I-34013, Trieste, Italy}}%

\begin{document}

\begin{abstract}
We consider a lattice gauge theory at finite temperature in ($d$+1) dimensions
with the Wilson action and different couplings $\beta_t$ and $\beta_s$ for
timelike and spacelike plaquettes. By using  the character expansion and
Schwinger-Dyson type equations we construct, order by order in $\beta_s$,
an effective action for the Polyakov loops which is exact to all orders in
$\beta_t$. As an example we construct the first non-trivial order in $\beta_s$
for the (3+1) dimensional SU(2) model and use this effective action to extract
the deconfinement temperature of the model.
\end{abstract}

\maketitle

\section{INTRODUCTION}
All the relevant properties of the deconfinement 
transition in finite temperature pure Lattice Gauge Theories (LGT)
can be described
by a suitable effective action for the order parameter, the
Polyakov loop~\cite{sy}. The construction of this effective action necessarily
involves some approximations. It is of crucial importance to choose these
approximations so as to obtain effective actions 
simple enough to be easily studied 
(either with exact solutions, or with some mean field like technique) 
and, at the same time, rich enough
to keep track of the whole complexity of the
original gauge theory. 
In the past years this problem was addressed with several different 
approaches (see~\cite{bcdp} for references and discussion). 
However a common feature of all these approaches was that 
the effective actions were always constructed 
neglecting the spacelike part of the
action. As a consequence it was impossible to reach
a consistent continuum limit for the critical temperature. 

The aim of this contribution is to show that it is possible
to avoid such a drastic approximation.
We shall discuss a general framework which 
allows one to construct
improved effective actions which take into account perturbatively (order by
order in the spacelike coupling $\beta_s$)  the spacelike
part of the original gauge action and are {\sl exact 
to all orders in the timelike coupling}.

Our approach is valid for any gauge group $G$ and for any
choice of lattice regularization of the gauge action (Wilson, mixed, heat kernel
actions\ldots). Moreover  it can be
extended, in principle, to all orders in $\beta_s$.

We shall only outline here the general strategy  and show, 
as an example, some results obtained by taking into account the first order 
contribution in $\beta_s$ in the case of the $SU(2)$ gauge model in (3+1) 
dimensions. Much more details and a complete survey of our approach can be found
in~\cite{bcdp}.

\section{CONSTRUCTION OF THE EFFECTIVE ACTION}
Since we  treat in a different way the spacelike and timelike parts of the
action,
we are compelled to use two different couplings $\beta_s$ and $\beta_t$.
We shall denote with $\rho^2\equiv\beta_t/\beta_s$ the asymmetry parameter, with
$N_t$ ($N_s$) the size of the lattice in the timelike (spacelike) direction and
with $d$ the number of spacelike dimensions.

The first step is to expand in characters  both the timelike
and the spacelike part of the action. The expansion of the spacelike part
is truncated at the chosen order in $\beta_s$;
the timelike part is kept exact to all orders. 

For the contribution due to the trivial representation term 
in the spacelike expansion
(the ``zeroth order'' approximation in $\beta_s$)
the integration over the spacelike degrees of freedom is
straightforward. The resulting effective action is that of a $d$ dimensional
spin model (the spins being the Polyakov loops of the original model) 
with next neighbour interactions only.

The terms of higher order $\beta_s$  
give rise to  non trivial interactions 
among several Polyakov loops. For instance at the order $\beta_s^2$
the interaction involves the four Polyakov loops around a spacelike plaquette.
The explicit form of these interactions  can be written in terms of
rather
complicated group integrals, which can be solved by means of suitable 
Schwinger Dyson (SD) equations. The use of these SD equations is crucial for
our whole construction since for these integrals
(in which the Polyakov loops are kept as free variables) 
the usual techniques, developed for ordinary
strong coupling expansions, are useless. We shall discuss in detail,
in the next section, a simple example.

\subsection{Schwinger-Dyson equations}
As an example, let us study the integral which appears in the discussion
of the  term due to the adjoint representation in the character expansion
of the spacelike action for the SU(2) model:
\eq
\label{new1}
\int DU U_{\alpha\beta} U^\dagger_{\gamma\delta} \chi_j(UP_{\vx + i}U^\dagger
P^\dagger_\vx )
\equiv\delta_{\alpha\delta} \delta_{\gamma\beta} \cj_{\alpha\beta} .
\en
First of all,
it follows from (\ref{new1}) that the non vanishing integrals in the
l.h.s. depend only on $|U_{\alpha\beta}|^2$, and hence that
$\cj_{11} = \cj_{22}$ and $\cj_{12} =\cj_{21}$.

To compute the matrix elements $\cj_{\alpha\beta}$, 
we note that the matrix $\cj$ can be expressed in terms of the integral
\eqa
\label{new2}
K^{(j)}(\theta_\vx,\theta_{\vx+i})& =&
\int DU \chi_j(U P_2 U^\dagger P^\dagger_1) \nonumber \\
&=&{1\over d_j} \chi_j(P_2)\chi_j(P^\dagger_1)
\ena
through a system of two linear Schwinger--Dyson-like equations.
Indeed,
considering the integral $\int DU U_{\alpha\beta}$ $U^\dagger_{\beta\alpha}$
$\chi_j(UP_{\vx + i}U^\dagger P^\dagger_\vx)$ we easily find that
\eq
\label{newrel1}
\cj_{11} + \cj_{12}=K^{(j)} .
\en
To construct a second independent equation, let us consider the integral
\eq
\label{newint}
\int DU \chi_\um(UP_{\vx + i}U^\dagger P^\dagger_\vx)
\chi_j(UP_{\vx + i}U^\dagger P^\dagger_\vx).
\en
On one hand we can write the character $\chi_\um$ explicitly as a trace and
express the integral in terms of the $\cj_{\alpha\beta}$ by using
eq.(\ref{new1}). On the other hand, the integral (\ref{newint}) can be
written in terms of $K^{(j)}$ functions by using the basic SU(2)
Clebsch-Gordan relation: $\chi_{1\over 2} \chi_j$ $= \chi_{j+{1\over 2}} +
\chi_{j-{1\over 2}}$. The resulting equation is:
\eqa
\label{newrel2}
2 \cos(\theta_{\vx + i} - \theta_\vx) \cj_{11} &+&
2\cos(\theta_{\vx + i} + \theta_\vx) \cj_{12} =\nonumber\\
K^{(j-{1\over 2})} &+& K^{(j+{1\over 2})}
\ena
where $\{e^{i\theta_{\vec x}}, e^{-i\theta_{\vec x}}\}$
 are the eigenvalues of the Polyakov loop $P_{\vec x}$. 
Eq.s (\ref{newrel1}) and (\ref{newrel2}) form a set of two linear equations
in the unknowns $\cj_{11}$ and $\cj_{12}$ whose solution is:
\eqa
\label{newsol}
\cj_{11} & = & {K^{(j-{1\over 2})} - 2 \cos(\theta_{\vx + i} + \theta_\vx)
K^{(j)} + K^{(j+{1\over 2})} \over 4 \sin \theta_{\vx + i} \sin \theta_\vx}
\nonumber\\
\cj_{12} & = & -{K^{(j-{1\over 2})} - 2 \cos(\theta_{\vx + i} - \theta_\vx)
K^{(j)} + K^{(j+{1\over 2})} \over 4 \sin \theta_{\vx + i} \sin \theta_\vx}
\nonumber
\ena

This solves the problem.

\section{DISCUSSION OF THE RESULTS}
The effective action obtained in the previous section describes a $d$
dimensional  spin model with complicated interactions and
cannot be solved exactly. However several features of the model can be figured
out rather easily. In particular, the deconfinement temperature can  be 
estimate by using a mean field approximation.
The results can then be used in two ways. 
\subsection{Asymmetric lattices}

As $\rho$ varies we have
different, but equivalent, lattice regularization of the same model. This
equivalence however implies, at the quantum level, in the (3+1) dimensional 
case a non trivial relation between the couplings. This problem was studied in
the weak coupling limit  by F. Karsch~\cite{k81} 
who found, in the $\rho\to\infty$ limit, the following relation:
\eq
\beta_t=\rho\,(\beta+\alpha_t^0)+\alpha_t^1
\label{rel3}
\en
where $\beta$ is the coupling of the equivalent
symmetric regularization. $\alpha_t^0$ and  $\alpha_t^1$ are group dependent
constants whose value in the SU(2) case is:
$\alpha^{0}_{t}=-0.27192$ and $\alpha^{1}_{t}=1/2$;

The $\rho$ dependence of our estimates for the critical
temperature shows a remarkable agreement
with the behaviour predicted by eq.(\ref{rel3})
(see tab.I.). As $N_t$ increases our estimates for
$\alpha^0_t$ and $\alpha^1_t$ cluster around the theoretical values 
of~\cite{k81}.  This agreement is
highly non trivial since $\alpha_t^0$ and $\alpha_t^1$ were obtained with a
{\sl weak coupling} calculation, while our effective action is the result of a
{\sl strong coupling} expansion. The reason of this success is very likely 
related to the fact that we have been able to sum to all orders in $\beta_t$
the timelike contribution of the effective action.
\begin{table}
\label{tab2}
\begin{center}
\begin{tabular}{c  c c }
\hline\hline
$ N_t$ &  $\alpha^0_t$ &  $\alpha^1_t$ \\
\hline
$2$ &    $-0.184$  & $0.414$\\
$3$ &   $-0.210$  & $0.375$\\
$4$ &   $-0.221$  & $0.373$\\
$5$ &   $-0.235$  & $0.372$\\
$6$ &   $-0.249$  & $0.389$\\
$8$ &   $-0.271$  & $0.413$\\
$16$ &  $-0.327$  & $0.508$\\
\hline
     &  $-0.27192$ & $0.50$\\
\hline\hline
\end{tabular}
\end{center}
\vskip 0.3cm
{\bf Tab. I}{\it~~ Values of of $\alpha_t^0$ and $\alpha_t^1$ 
 as functions
of $N_t$. The
theoretical values are reported, for comparison, in the last row of the 
table.}
\end{table}
\subsection{Scaling behaviour}
The second important test is the scaling behaviour of the deconfinement
temperature as a function of $N_t$. The $N_t$ dependence is predicted to be of
logarithmic type in (3+1) dimensions and the Montecarlo data confirm this
analysis. On the contrary all the effective actions obtained neglecting the
spacelike plaquettes predict a linear scaling (see Fig.1). This was in
past years one
of the major drawbacks of the standard effective action approach to the
deconfinement transition.  The inclusion of the first non 
trivial corrections due to the space-like plaquettes greatly improves
the scaling behaviour.
The values obtained with our effective action for the critical couplings
 are plotted in
Fig.1, where they are also compared with the Montecarlo results
(extracted from~\cite{fhk}).
\begin{figure}
\null\vskip -1cm\hskip 2cm
\epsfxsize = 7truecm
\epsffile{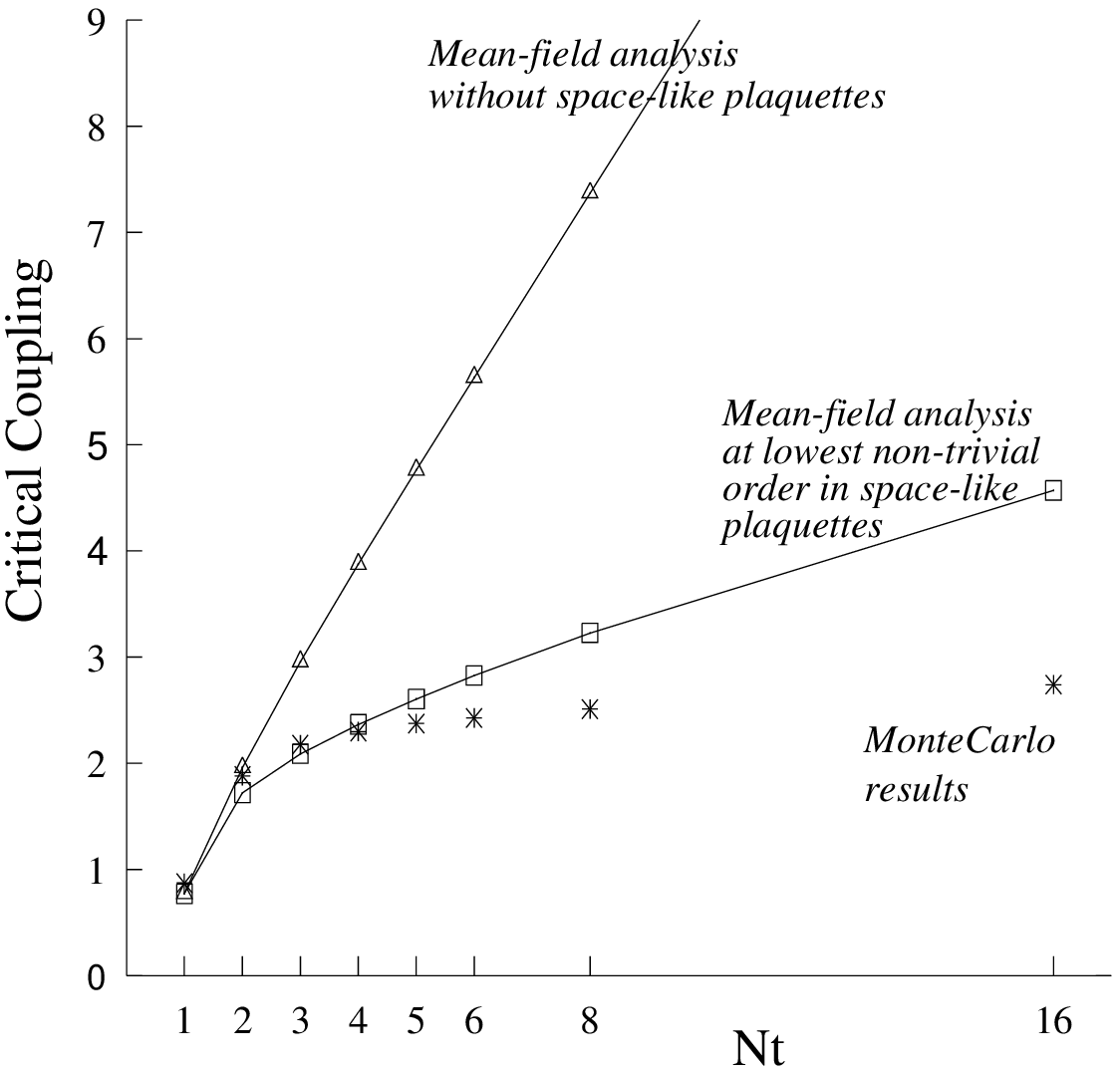}
\label{data}
\end{figure}
Indeed, if one compares our mean field estimates with the ones
of the Montecarlo simulations, it turns out that the discrepancy in
critical couplings is within 10\%  in the range
$2\leq N_t \leq 5 $. 
While the logarithmic scaling predicted by the renormalization group
is still beyond the present scheme, being related to non perturbative effects 
in $\beta_s$, it is reasonable to expect that higher order approximations
would lead to better and better numerical results at least for not too high
values of $N_t$.


\begin{thebibliography}{9}
\bibitem{sy} B.Svetitsky and L.Yaffe, Nucl. Phys. {\bf B210}
(1982) 423.
\bibitem{bcdp} M.Bill\'o, M.Caselle, A.D'Adda and S.Panzeri,
preprint hep-lat/9601020, Nucl. Phys. B, to appear.
\bibitem{k81} F.Karsch, \NP{B205} (1982) 285.
\bibitem{fhk} J.Fingberg, U.Heller and F.Karsch, \NP{B392} (1993) 493.
\end{thebibliography}
\end{document}